\documentclass[10pt,twocolumn,english,prd,showpacs,nofootinbib]{revtex4}
\usepackage[latin9]{inputenc}
\usepackage{graphicx}
\usepackage{amssymb}
\providecommand{\tabularnewline}{\\}

\usepackage{babel}

\begin{document}

\title{Fingerprinting Dark Energy II: weak lensing and galaxy clustering
tests.}

\date{\today}

\author{Domenico Sapone}

\email{domenico.sapone@uam.es}

\affiliation{Departamento de F\'isica Te\'orica and Instituto de F\'isica Te\'orica, \\
 Universidad Aut\'onoma de Madrid IFT-UAM/CSIC,\\
 $28049$ Cantoblanco, Madrid, Spain}

\author{Martin Kunz}

\email{Martin.Kunz@unige.ch}

\affiliation{D\'epartement de Physique Th\'eorique, Universit\'e de Gen\`eve, 24 quai
Ernest Ansermet, CH--1211 Gen\'eve 4, Switzerland}

\affiliation{Institut d'Astrophysique Spatiale, Universit\'e Paris-Sud XI, Orsay
91405, France}

\affiliation{Astronomy Centre, University of Sussex, Falmer, Brighton BN1 9QH,
UK}

\author{Luca Amendola}

\email{L.Amendola@thphys.uni-heidelberg.de}

\affiliation{University of Heidelberg, Philosophenweg 16, 69120 Heidelberg, Germany
and INAF/Rome}
\begin{abstract}
The characterization of dark energy is a central task of cosmology.
To go beyond a cosmological constant, we need to introduce at least
an equation of state and a sound speed and consider observational
tests that involve perturbations. If dark energy is not completely
homogeneous on observable scales then the Poisson equation is modified
and dark matter clustering is directly affected. One can then search
for observational effects of dark energy clustering using dark matter
as a probe. In this paper we exploit an analytical approximate solution
of the perturbation equations in a general dark energy cosmology to
analyze the performance of next-decade large scale surveys in constraining
equation of state and sound speed. We find that tomographic weak lensing
and galaxy redshift surveys can constrain the sound speed of the dark
energy only if the latter is small, of the order of $c_{s}\lesssim0.01$
(in units of $c$). For larger sound speeds the error grows to 100\%
and more. We conclude that large scale structure observations contain
very little information about the perturbations in canonical scalar
field models with a sound speed of unity. Nevertheless, they are able
to detect the presence of {}``cold'' dark energy, i.e. a dark energy
with non-relativistic speed of sound. 
\end{abstract}


\maketitle

\section{Introduction}

\global\long\global\long\global\long\def\cs{c_{s}^{2}}
 \global\long\global\long\global\long\def\rmd{{\rm d}}
 \global\long\global\long\global\long\def\ca{c_{a}^{2}}
 \global\long\global\long\global\long\def\de{\mathrm{DE}}
 \global\long\global\long\global\long\def\lkr{{\left(k\right)}}
 Even though there is ample observational evidence that the expansion
rate of the Universe is accelerating \cite{sn1,sn2}, there is still
no convincing theoretical model that can explain the observations.
The simplest model in agreement with the data is the cosmological
constant. However, it suffers from size and coincidence problems.
In addition, it is likely that an earlier phase of accelerated expansion
happened that was not due to a cosmological constant \cite{iklf}.
This period, inflation, is normally modeled as being due to a scalar
field. The same basic mechanism can explain the dark energy, and is
often called Quintessence in this context \cite{cald}. Scalar field
dark energy is often employed to model a general equation of state
$w(z)$, although one can also build a perfect fluid with exactly
the same first order perturbations \cite{ksphantom}.

Over the last few years it has become clear that $w$ alone is not
sufficient to distinguish between different models explaining the
dark energy, and that the perturbations induced by the dark energy
constitute a complementary probe \cite{mfb,abp,lss,km,ks2,hs,aks}.
The contribution of dark energy fluctuations to the perturbation dynamics
is generally small and can be very difficult to measure. In a recent
paper some of us derived theoretical approximations to the perturbations
generated in scalar field dark energy models \cite{sk}. In this paper,
we will use these expressions to investigate how two cosmological
probes that will be of great importance in the future, weak lensing
(WL) and the full galaxy power spectrum, can be used to measure the
typical features that appear in such models, specifically the existence
of a sound horizon. 
Exploiting the advantage of analytical approximations,
we will try to isolate in some detail where the signal comes from.

Anticipating the conclusion, we will show that even a full sky, tomographic,
deep redshift and imaging survey cannot constrain significantly the
sound speed of the dark energy unless it is smaller than, roughly,
0.01$c$%
\footnote{Here $c$ denotes the speed of light, and in the rest of the paper
we will use units so that $c=1$.%
}. Small values of $c_{s}$ are not compatible with a scalar field
description with standard kinetic term and require extensions like
k-essence \cite{kess} but are in general not forbidden: we will show
that even with $c_{s}=0.01$, dark energy, although {}``cold'',
remains much less clustered than dark matter and will not develop
non-linearities by the present time. The special case of $c_{s}=0$
has some additional attractive features, in that such models can cross
$w=-1$ \cite{ksphantom,vernizzi} and can act as unified models of
dark energy and dark matter \cite{gklp} (but are also subject to
the {}``dark degeneracy'' \cite{dark_degen}). For such extreme
cases however dark energy becomes non linear along with dark matter
and appropriate non-linear corrections are needed, especially for
the weak lensing probe.

The sound speed of the dark energy could be measured also by other
effects, like the Integrated Sachs-Wolfe (ISW) effect, either on the
CMB or cross-correlated with large-scale structure. In \cite{wellew}
it has been shown that the CMB ISW tail can put a lower limit to $\cs$,
while \cite{beandore} finds that $c_{s}\approx0.01$ is actually
favored by the WMAP 3-year data. The ISW is complementary to the probes
we consider here since it is sensitive to perturbations on larger
scales.

As in \cite{sk} we set the anisotropic stress of the dark energy
to zero, which is appropriate for scalar field dark energy and in
general for standard gravity models. In order to investigate more
general cases, for example modifications of General Relativity, a
non-zero anisotropic stress would have to be taken into account \cite{ks2},
but this is left for future work.

In detail, the paper is organized as follows. We begin by discussing
the main results obtained in \cite{sk}. We then use these results
to study the impact of the scalar field dark energy perturbations
on the weak lensing signal and on probes of galaxy clustering, paying
special attention to whether the presence of a sound horizon can be
detected (by measuring the sound speed $\cs$). We also discuss which
aspect of these probes are most sensitive to the perturbations.

We finally investigate constraints on the integrated deviation from
the Poisson equation, which we call $W$, which may be useful to quantify
the power of probes beyond the conventional, $w$-based Figure of
Merit used by the dark energy task force \cite{detf}.

\section{Setting the scene}

In this section we define our notation and present a short discussion
of the perturbation equations and the approximate analytical solution
of \cite{sk}. We refer the reader to that paper (and citations therein)
for more details.

\subsection{Definitions}

In the following, the dots will refer to the derivatives with respect
to the conformal time $\tau$ which is related to the universal time
$t$ by $dt=a\left(t\right)d\tau$. The physical Hubble parameter
that we will consider here is: \begin{equation}
H^{2}=\left(\frac{da}{adt}\right)^{2}=H_{0}^{2}\left[\Omega_{m,0}a^{-3}+\left(1-\Omega_{m,0}\right)g(a)\right]\label{eq:hubble}\end{equation}
 where $g(a)=\exp\left[-3\int\frac{1+w(a)}{a}da\right]$ and the subscript
0 denotes the present epoch. This expression implies that we limit
ourselves to a flat universe which is filled with matter (designated
by a subscript $m$) and a general fluid with average pressure $p=w\rho$.
In general the equation of state parameter $w$ is a function of time,
but we will take it to be constant later on.

We will consider linear perturbations about this spatially-flat background
model, defined by the line element: \begin{equation}
\rmd s^{2}=a^{2}\left[-\left(1+2\psi\right)\rmd\tau^{2}+\left(1-2\phi\right)\rmd x_{i}\rmd x^{i}\right].\label{pert_newton_ds}\end{equation}
 As is apparent from the line element, we use the conformal Newtonian
(longitudinal) gauge and retain only scalar perturbations. The perturbation
equations for a dark energy fluid with sound speed $c_{s}$ and parameter
of state $w$ are: \begin{eqnarray}
\delta' & = & -\frac{V}{Ha^{2}}\left(1+\frac{9a^{2}H^{2}\left(\cs-w\right)}{k^{2}}\right)-\frac{3}{a}\left(\cs-w\right)\delta\nonumber \\
 & + & 3\left(1+w\right)\phi',\label{deltap}\\
V' & = & -\left(1-3\cs\right)\frac{V}{a}+\frac{k^{2}\cs}{Ha^{2}}\delta+(1+w)\frac{k^{2}}{Ha^{2}}\psi \label{vp}\end{eqnarray}
 where $V=(1+w)(ik_{i}v^{i})$ is a measure of the velocity perturbation,
$\delta$ is the density contrast, and the prime here means the derivative
with respect to the scale factor $a$. We parameterize the pressure
perturbation as: \begin{equation}
\delta p=\cs\rho\delta+\frac{3aH\left(\cs-\ca\right)}{k^{2}}\rho V\,,\end{equation}
 where $\ca=w-\frac{\dot{w}}{3H\left(1+w\right)}$ is the adiabatic
sound speed and $\cs$ is the sound speed in the rest-frame of the
dark energy fluid. As already mentioned, we assume that the parameter
of equation of state $w$ stays constant so that $\ca=w$.

We also assume a vanishing anisotropic stress $\sigma=0$ as is the
case for Quintessence and K-essence models, and therefore we have
$\psi=\phi$. Finally, the gravitational potential can be found with
the help of the Einstein equations, \begin{equation}
k^{2}\phi=-4\pi Ga^{2}\rho\left(\delta+\frac{3aH}{k^{2}}V\right).\label{eq:psi}\end{equation}

\subsection{Analytical solutions for dark energy perturbations}

If the universe is perfectly matter dominated, then $k^{2}\phi$ is
a constant. In \cite{sk} we used this result to derive analytical
solutions to the perturbation equations for dark energy in different
limits: 
\begin{itemize}
\item \textbf{scales larger than the sound horizon.} The modes larger than
the sound horizon are found for $k\ll aH/c_{s}$. In this case, we
neglect all terms containing the sound speed in Eq.~(\ref{vp}) and
then in Eq.~(\ref{deltap}), effectively setting $\cs=0$. Neglecting
a decaying solution for which $V\propto1/a$ we find: \begin{eqnarray}
\delta\left(a\right) & = & (1+w)\delta_{in}\left(\frac{a}{1-3w}+\frac{3H_{0}^{2}\Omega_{m,0}}{k^{2}}\right),\label{DEsolsmallsubfull}\\
V\left(a\right) & = & -\left(1+w\right)H_{0}\sqrt{\Omega_{m,0}}\delta_{in}a^{1/2}.\label{VDEsolsmallsubfull}\end{eqnarray}

\item \textbf{Scales smaller than the sound horizon.} For modes smaller
than the sound horizon, $k\gg aH/c_{s}$, we consider only the terms
containing $k^{2}$. We find that the density perturbations become
constant and that the velocity perturbations decay:
\begin{eqnarray}
\delta\left(a\right) & = & \frac{3}{2}\left(1+w\right)\frac{H_{0}^{2}\Omega_{m,0}}{\cs k^{2}}\delta_{in},\\
V\left(a\right) & = & -\frac{9}{2}\left(1+w\right)\left(\cs-w\right)\frac{H_{0}^{3}\Omega_{m,0}^{3/2}}{\sqrt{a}\cs k^{2}}\delta_{in}.
\end{eqnarray}

\end{itemize}
In these equations, the factor $\delta_{in}$ sets the overall scale
of the perturbations, chosen so that the matter perturbations are
$\delta_{m}=\delta_{in}a$ on sub-horizon scales. In addition, the
dark energy perturbations depend on their present density fraction
(equal to $1-\Omega_{m,0}$ since the universe is taken to be flat)
but only through the combination $\Omega_{m,0}h^{2}$, the equation
of state parameter $w$ and the sound speed $\cs$. They are a function
of both scale $k$ and scale factor $a$.

Let us quickly compare dark energy and dark matter fluctuations. For
large scales the dominant term in Eq. (\ref{DEsolsmallsubfull}) is
the term containing $k$. Dark matter has a similar solution except
that $w_{DM}=0$; the ratio of the squares (i.e. the power spectrum
ratio) for large scales then will be : \begin{equation}
\left(\frac{\delta_{DE}}{\delta_{m}}\right)^{2}=\left(1+w\right)^{2}\le\frac{1}{25}\end{equation}
 if $w\le-0.8$. However, for small scales (but still larger than
$c_{s}/aH$), the first term in Eq. (\ref{DEsolsmallsubfull}) dominates;
in this case the ratio is: \begin{equation}
\left(\frac{\delta_{DE}}{\delta_{m}}\right)^{2}=\left(\frac{1+w}{1-3w}\right)^{2}\le\frac{1}{17^{2}}\simeq0.0035.\end{equation}
 The dark energy perturbations are thus always suppressed relative
to the dark matter perturbations, even for a very low sound speed
(note that the solution Eq. (\ref{DEsolsmallsubfull}) was obtained
assuming $\cs=0$ ).

For the forecasts of this paper, it is important to know whether we
have to worry about non-linear effects: the non-linear clustering
of dark energy models is still badly understood and may depend on
the precise action of the model, and it is not clear what corrections
we would need to apply. Outside the sound horizon, the dark energy
perturbation track those in the dark matter by the factors given above.
These are roughly upper limits as $w=-0.8$ is at the upper limit
of what is still allowed, and values of $w$ closer to $-1$ will
lead to smaller dark energy perturbations. If $c_{s}=0.01$ the sound
horizon at the present is of the order of $c_{s}(2/H_{0})\approx60~{\rm Mpc}/h$.
On these scales dark matter fluctuations are still linear and therefore
dark energy fluctuations are safely below the non-linearity threshold.
On sub-sound-horizon scales, dark energy perturbations stop growing
and become quickly negligible relative to the dark matter perturbations.
If $c_{s}$ is smaller than $0.001$ then the dark energy will behave
much like dark matter on the cluster scales and an appropriate non-linear
correction would be needed, especially on the small scales probed
by weak lensing (the small scale cut-off in the galaxy clustering
safely excludes non linear scales). Therefore we do not extend our
analysis to such small sound speeds; we include however for completeness
the extreme case $c_{s}=0$, for which we assume that we can apply
the same non-linear correction as for the dark matter -- whether this
is acceptable may in general depend on the precise action of the field.

\section{Impact of the sound speed on galaxy clustering and weak lensing}

In this section we discuss the effect of dark energy clustering on
the galaxy and weak lensing probes. We derive approximate analytical
expressions that show how the relevant quantities change with the
sound speed. This will help understanding qualitatively the numerical
results of the next section.

\subsection{The $Q$ parameter\label{sec:q}}

Matter domination was a necessary ingredient to derive the solutions
given above. However, dark energy comes to dominate eventually, and
then the potential starts to decay and the perturbations grow more
slowly or start to decrease.

It is difficult to capture this behavior accurately. A way around
this problem can be found by looking at the variable $Q(k,a)$ which
we introduced in \cite{aks} to describe the change of the gravitational
potential due to the dark energy perturbations. $Q$ is defined through
\begin{equation}
k^{2}\phi=-4\pi Ga^{2}Q\rho_{m}\left(\delta_{m}+\frac{3aH}{k^{2}}V_{m}\right).
\label{eq:Q_def}
\end{equation}
 If the dark energy or modification of gravity does not contribute
to the gravitational potential (for example if the dark energy is
a cosmological constant) then $Q=1$. Otherwise $Q$ will deviate
from unity, and in general it is a function of both scale and time.

Introducing the comoving density perturbation $\Delta\equiv\delta+3aHV/k^{2}$,
we can write \begin{equation}
k^{2}\phi=-4\pi Ga^{2}(\rho_{m}\Delta_{m}+\rho_{DE}\Delta_{DE}).\label{eq:Q_def-2}\end{equation}
 We neglect radiation in this paper as we are interested in observational
tests at late times where it is subdominant. Then $Q$ is defined
as \begin{equation}
Q-1=\frac{\rho_{\de}\Delta_{\de}}{\rho_{m}\Delta_{m}}.\label{eq:Qdef3}\end{equation}
 Using simply the solution for the perturbations during matter domination,
as discussed in the previous paragraph, we find that the resulting
expression for $Q$ is surprisingly accurate even at late times. The
reason is that both fluids, dark energy and matter, respond similarly
to the change in the expansion rate so that most of the deviations
cancel. We find that the sub-soundhorizon expression below is accurate
at the percent level, while on larger scales there are deviations
of about 10 to 20\% by today (depending on $w$). The latter can be
corrected {}``by hand'' in order to obtain a more precise formula,
but the expressions are sufficiently accurate for our purposes and
we keep them as they are.

As we have set the anisotropic stress to zero, the perturbations are
fully described by $Q$. In \cite{sk} we provided the following explicit
expression for the $Q\left(k,a\right)$ which captures the behavior
for both limits (above and below the sound horizon): \begin{equation}
Q(k,a)=1+\frac{1-\Omega_{m,0}}{\Omega_{m,0}}\frac{(1+w)a^{-3w}}{1-3w+\frac{2}{3}\nu(a)^{2}}.\label{eq:qtot}\end{equation}
 Here we used $\nu(a)^{2}=k^{2}\cs a/\left(\Omega_{m,0}H_{0}^{2}\right)$
which we defined through $c_{s}k\equiv\nu aH$ so that $\nu$ counts
how deep a mode is inside the sound horizon. We see that during matter
domination (used in the first expression for $\nu$) a $k$ mode moves
ever more deeply inside the sound horizon as $a$ grows, which is
of course no surprise.

From Eq.~(\ref{eq:Qdef3}) we see that the deviation of $Q$ from
$1$ depends on the ratio of the energy density of the dark matter
and the dark energy, and on the relative amount of perturbations.
The former scales as $a^{-3w}$ for a constant $w$ and the latter
behaves as discussed in the last section. For $\Omega_{m,0}=0.25$
and $w=-0.8$ we have that \begin{equation}
Q-1\approx\frac{3}{17}a^{2.4}\simeq0.18a^{2.4}\end{equation}
 on scales that are larger than the sound horizon, $\nu\approx0$.
This is not a negligible deviation today, but it decreases rapidly
as we move into the past, as the dark energy becomes less important.%
\footnote{For this reason, early dark energy models can have a much stronger
impact \cite{deput}.%
} As a scale enters the sound horizon, $Q-1$ grows with one power
of the scale factor slower (since $\delta_{DE}$ stops growing), suppressing
the final deviation roughly by the ratio of horizon size to the scale
of interest. In the observable range, $(k/H_{0})^{2}\approx10^{2}-10^{4}$.
Therefore if $c_{s}\approx1$, $Q\to1$ and the dependence on $c_{s}$
is lost. This shows that $Q$ is sensitive to $c_{s}$ only for small
values, $c_{s}^{2}\lesssim10^{-2}$.

We can characterize the dependence of $Q$ on the main perturbation
parameter $\cs$ by looking at its derivative, a key quantity for
Fisher matrix forecasts: \begin{equation}
\frac{\partial\log Q}{\partial\log\cs}=-\frac{x}{\left(1+x\right)}\frac{Q-1}{Q}.\label{eq:Qdercs}\end{equation}
 where $x=\frac{2}{3}\nu(a)^{2}/(1-3w)\simeq0.2\nu(a)^{2}$ (with
the last expression being for $w=-0.8$). For the values we are interested
here, this derivative has a peak just inside the sound
horizon, the exact position is
\begin{eqnarray}
k_{max}=\frac{H_{0}}{c_s}\sqrt{\frac{3}{2}\Omega_{m,0}\left(1-3w\right)\left(1+z\right)}\nonumber \\
\left[1+\frac{1+w}{1-3w}\frac{1-\Omega_{m,0}}{\Omega_{m,0}}\left(1+z\right)^{3w}\right]^{1/4} .
\end{eqnarray}
Today the sound horizon is given by $c_{s}\approx H_{0}/k$. It lies in the observable
range of $k$ for sound speeds of the order of $c_{s}\approx0.01-0.001$. We plot
the derivative as the red curve in Fig.~\ref{fig:der-G-Q-b-P0-ink}, there and from
Eq.~(\ref{eq:Qdercs}) we can see that the shape is basically proportional to 
$-x/(1+x)^2$. The derivative with respect to $\cs$ (instead of $\log\cs$) contains an additional
factor of $1/\cs$ which means that it will be boosted by several orders of magnitude
for small sound speeds.

We will later forecast how well the deviation of $Q$ from $1$ due
to the dark energy perturbations can be measured by future cosmological
surveys.

\subsection{The growth rate and the $\gamma$ parameter\label{sec:gamma}}

In the $\Lambda$CDM model of cosmology, the dark matter perturbations
on sub-horizon scales grow linearly with the scale factor $a$ during
matter domination. During radiation domination they grow logarithmically,
and also at late times, when dark energy starts to dominate, their
growth is suppressed. It is well known that in $\Lambda$CDM the growth
factor can be expressed as: \begin{equation}
G\left(a\right)\equiv\frac{\delta_{m}(a)}{\delta_{m}(a_{0})}=\exp\Big\{\int_{0}^{a}\frac{\Omega_{m}\left(a'\right)^{\gamma}}{a'}{\rm d}a'\Big\}\end{equation}
 where $\gamma\sim0.545$ is called the growth index. There are two
ways to influence the growth factor: firstly at background level,
with a different Hubble expansion. Secondly at perturbation level:
if dark energy clusters then the gravitational potential changes because
of the Poisson equation, and this will also affect the growth rate
of dark matter. All these effects can be included in the growth index
$\gamma$ and we therefore expect that $\gamma$ is a function of
$w$ and $\cs$ (or equivalently of $w$ and $Q$).

According to \cite{lica}, the growth index depends on dark energy
perturbations (through $Q$) as \begin{equation}
\gamma=\frac{3\left(1-w-A\left(Q\right)\right)}{5-6w}\label{eq:gamma-Q}\end{equation}
 where \begin{equation}
A\left(Q\right)=\frac{Q-1}{1-\Omega_{m}\left(a\right)}.\label{eq:A-Q}\end{equation}
 However, dark energy perturbations (hence $Q$) are difficult to
measure at least if the dark energy has a very low sound speed. The
growth index factor seems to be a more promising parameter and several
experiments are planned to measure this quantity. Furthermore, we
can invert Eq. (\ref{eq:gamma-Q}) and ask the question: in the absence
of anisotropic stress which $Q$ is needed to generate a given $\gamma$?
Using Eqs. (\ref{eq:gamma-Q}) and (\ref{eq:A-Q}) we have: \begin{equation}
Q-1=\left[1-\Omega_{m}\left(a\right)\right]\left[1-w-\frac{1}{3}\left(5-6w\right)\gamma\right].\label{eq:Q-gamma}\end{equation}
 The last equation is worth another look: let us assume we measure
the growth index $\gamma=6/11$ (the value usually associated to the
cosmological constant) then Eq. (\ref{eq:Q-gamma}) becomes: \begin{equation}
Q-1=\frac{1}{11}\left(1+w\right)\left[1-\Omega_{m}\left(a\right)\right].\label{eq:Q-gamma-1}\end{equation}
 If we are dealing with the cosmological constant then we have the
expected result of $Q-1=0$. However, if $w\neq-1$ then we can evaluate
the corresponding value of $Q$; for instance, if $w=-0.8$ then we
have $Q-1\sim10^{-2}$, which is a positive number. However, this
value is fairly big for dark energy perturbations, see \cite{sk}.
Moreover, for non-phantom models the quantity $Q-1$ is always positive
due to the relative increase of dark energy perturbations and the
$(1+w)$ factor; let us remind the reader that if the anisotropic
stress is set to zero then $Q$ fully describes the evolution of dark
energy perturbations; so, we can think to rephrase the sentence as:
if $Q-1>0$ which $\gamma$ can we reach? Setting Eq. (\ref{eq:Q-gamma})
positive we have \begin{equation}
\gamma<\frac{3\left(1-w\right)}{5-6w}\label{eq:gamma-limit}\end{equation}
 which sets a sort of upper bound to $\gamma$; moreover, this is
the limit we have if dark energy perturbations are set to zero (which
can be seen directly from Eq. (\ref{eq:gamma-Q})).

In general we can say that if the anisotropic stress is zero then
dark energy perturbations always decrease the value of the growth
index. If we want to allow for perturbations in the dark energy sector
while increasing at the same time the growth index then we need a
non zero anisotropic contribution; this is the case for the DGP model
(treated as an effective dark energy perturbations, see \cite{ks2})
where $\gamma\sim0.68$.

Let us also consider the derivative of $\log G$ with respect $\cs$
to gain an idea of how strongly the growth factor depends on sound
speed, and thus on the characteristics of the dark energy perturbations:
\begin{equation}
\frac{\partial\log G(a_{1})}{\partial\log\cs}=c_{s}^{2}\int_{a_{0}}^{a_{1}}{\frac{1}{a}\frac{\partial\gamma}{\partial\cs}\log\Omega_{m}\left(a\right)\Omega_{m}\left(a\right)^{\gamma}{\rm d}a}.\label{eq:dG-dcs}\end{equation}
 We can see more clearly the dependence on the parameters by noting
that the quantity $\log\Omega_{m}\left(a\right)\Omega_{m}\left(a\right)^{\gamma}$
can be approximated as $\Omega_{m}\left(a\right)-1$ for $\Omega_{m}$
close to unity (we use however the exact expressions in the calculations
below). Eq.(\ref{eq:dG-dcs}) then reads: \begin{eqnarray}
\frac{\partial\log G(a_{1})}{\partial\log\cs} & = & \frac{3c_{s}^{2}}{5-6w}\int_{a_{0}}^{a_{1}}{\frac{\partial Q}{\partial\cs}\frac{{\rm d}a}{a}}\nonumber \\
 & = & -\frac{3}{5-6w}\int_{a_{0}}^{a_{1}}{\frac{Q-1}{1+b~a}{\rm d}a}\nonumber \\
 & = & -\frac{3}{5-6w}\frac{1+w}{1-3w}\frac{1-\Omega_{m,0}}{\Omega_{m,0}}\times\nonumber \\
 & \times & b\int_{a_{0}}^{a_{1}}{\frac{a^{-3w}}{\left(1+b~a\right)^{2}}{\rm d}a}\end{eqnarray}
 where $b=x/a\approx0.2\nu^{2}/a$ is a constant during matter domination
(the value of $x$ today). We first consider two different limits 
\begin{itemize}
\item $b\gg1$ (sub-sound horizon regime): \begin{eqnarray}
\frac{\partial\log G}{\partial\log\cs} & = & -\frac{3}{5-6w}\frac{1+w}{1-3w}\frac{1-\Omega_{m,0}}{\Omega_{m,0}}\times\nonumber \\
 & \times & \frac{1}{b}\int_{a_{0}}^{a_{1}}{a^{-3w-2}{\rm d}a}=\nonumber \\
 & = & \frac{3}{5-6w}\frac{Q-Q_{0}}{1+3w}\label{eq:Gdercs}\end{eqnarray}

\item $b\ll1$ (super-sound horizon regime): \begin{eqnarray}
\frac{\partial\log G}{\partial\log\cs} & = & -\frac{3}{5-6w}\frac{1+w}{1-3w}\frac{1-\Omega_{m,0}}{\Omega_{m,0}}\times\nonumber \\
 & \times & b\int_{a_{0}}^{a_{1}}{a^{-3w}{\rm d}a}=\nonumber \\
 & = & -\frac{3b}{5-6w}\frac{\left(Q-1\right)a-\left(Q_{0}-1\right)a_{0}}{1-3w}\label{eq:Gdercs1}\end{eqnarray}

\end{itemize}
Both of these derivatives, like the one of $Q$ discussed further
above, have a strong dependence on the sound speed. As before, the peak location is near
the sound horizon, and again the derivative with respect to $\cs$ adds an additional 
$1/\cs$ factor so that small sound speeds will lead to much larger values of the derivative.

The numerical factor (including the $w$ dependence) 
is $-0.22$ and $-0.09$ for $w=-0.8$, respectively. We also notice that the first limit does
not depend on $Q-1$ like the derivative $Q$, but on $Q-Q_{0}$ as
it is an integral. The growth factor is thus not directly probing
the deviation of $Q$ from unity, but rather how $Q$ evolves over
time. This is similar in the second limit, except that we probe the
evolution of $(Q-1)x$.

We can find a unified formula which accounts for both regimes: \begin{widetext}
\begin{equation}
\frac{\partial\log G}{\partial\log\cs}=-\frac{3}{5-6w}\frac{\left(Q-Q_{0}\right)\left[\left(Q-1\right)x-\left(Q_{0}-1\right)x_{0}\right]}{\left(1-3w\right)\left(Q-Q_{0}\right)-\left(1+3w\right)\left[\left(Q-1\right)x-\left(Q_{0}-1\right)x_{0}\right]}\label{eq:Gdercs-union}\end{equation}
 \end{widetext} where $Q_{0}$ and $x_{0}$ are evaluated at $a=a_{0}=1$.

The derivative is shown as the blue dashed line in Fig.~\ref{fig:der-G-Q-b-P0-ink}.
Just as the $Q$ derivative, the $G$ derivative peaks close to the
sound horizon, but is smaller than the one of $Q$ at this redshift.
It grows however with growing redshift due to its integral nature,
while the other derivatives decrease.

\subsection{Shape of the dark matter power spectrum\label{sec:pk}}

In order to quantify the impact of the sound speed on the matter power
spectrum we proceed in two ways: we first use the CAMB output \cite{camb}
(which contains the full information on dark energy perturbations)
and then we consider the analytic expression from Eisenstein \& Hu
\cite{ehu} (which does not include dark energy perturbations, i.e.
does not include $c_{s}$).

As anticipated, we find here that the impact of the derivative of
the matter power spectrum with respect the sound speed on the final
errors is only relevant if high values of $\cs$ are considered; for
instance if $\cs=1$ then the errors on the sound speed increase by
about $10$ times when the analytic formula in Ref. \cite{ehu} is
considered instead of the CAMB output. By decreasing the sound speed,
the results are less and less affected, and are completely unchanged
when $\cs=10^{-5}$. The reason is that for low values of the sound
speed other parameters, like the growth factor, start to be the dominant
source of information on $c_{s}^{2}$, see Fig.~\ref{fig-fisher-cs000001}.
Since for large $c_{s}$ our results show that the sound speed in
practically not measurable with the probes here considered, we disregard
this difference between the analytical fit and the CAMB output and
use the analytical fit whenever we find it convenient, in particular
in estimating the weak lensing Fisher matrix.

\begin{figure}
\centering \includegraphics[width=2.6in]{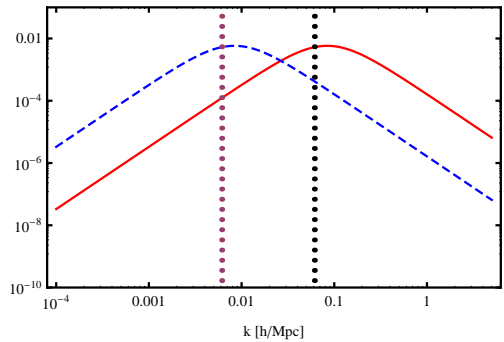}

\caption{The figure shows the derivatives of the matter power spectrum with
respect to $\log \cs$ for two different values of the sound speed:
$\cs=10^{-2}$ and $\cs=10^{-4}$, red solid and blue dashed line,
respectively. The vertical dotted lines give the scale of the
sound horizon for two different sound speeds, $\cs=10^{-4}$
(left) and $\cs=10^{-2}$ (right).}

\label{fig:derlogpkdlogcs} 
\end{figure}

In Fig.~\ref{fig:derlogpkdlogcs} we show the derivative of the logarithm
of the matter power spectrum (from CAMB) with respect to the logarithm
of sound speed. It is easy to understand this result: In Fig.~3 of
\cite{sk} we showed that the power spectrum decreases by a few percent
in a smooth step-like fashion at a scale just inside the sound horizon.
Here we see the derivative of this transition. Just like the step,
the peak of the derivative is situated a little bit inside the sound
horizon, as is the case for the other contributions as well.

\subsection{Redshift space distortions\label{sec:zdist}}

The velocity field of matter is directly sourced by the gradient of
the $\psi$ gravitational potential, and astronomers have been trying
to measure it for a long time \cite{flow1,flow2}. More recently constraints
on the velocity perturbations from redshift space distortions became
an important probe for dark energy phenomenology, since their combination
with weak lensing allows to disentangle the two potentials, which
in turn allows to put limits on the anisotropic stress of the dark
sector \cite{aks}. This is an important test for modifications of
gravity \cite{ks2}. Although there is no anisotropic stress in our
scenario, we are nonetheless interested in how redshift space distortions
are modified by the presence of dark energy perturbations.

The distortion induced by redshift can be expressed in linear theory
by the $\beta$ factor, related to the bias factor and the growth
rate via: \begin{equation}
\beta(z,k)=\frac{\Omega_{m}\left(z\right)^{\gamma(k,z)}}{b(z)}.
\label{eq:bias}
\end{equation}
The derivative of the redshift distortion parameter with respect to
the sound speed is: 
\begin{eqnarray}
 &  & \frac{\partial\log\left(1+\beta\mu^{2}\right)}{\partial\log\cs}\nonumber \\
 & = & -\frac{3}{5-6w}\frac{\beta\mu^{2}}{1+\beta\mu^{2}}\frac{x}{1+x}\left(Q-1\right).
 \label{eq:derbetadcs}
 \end{eqnarray}
 We see that the behavior versus $\cs$ is similar to the one
for the $Q$ derivative, so the same discussion applies. Once again,
the effect is maximized for small $c_{s}$.

\begin{figure}
\centering \includegraphics[width=2.6in]{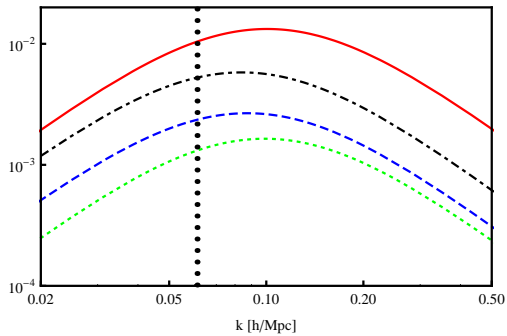}

\caption{The figure shows the derivatives with respect to $\log \cs$ of
the $Q$ parameter Eq.~(\ref{eq:Qdercs}) (red solid line), the growth
factor Eq.~(\ref{eq:Gdercs-union}) (blue dashed line), the redshift
distortion parameter Eq.~(\ref{eq:derbetadcs}) (green dotted line)
and the matter power spectrum $P_{0}(k)$ (black dot-dashed line).
The derivatives are evaluated at a $z=0.5$ except the one for
$P_{0}(k)$ which is at $z=0$; for all the derivatives the sound
speed is $\cs=10^{-4}$. The vertical dotted line gives the scale 
of the sound horizon for $\cs=10^{-4}$ at $z=0.5$.}

\label{fig:der-G-Q-b-P0-ink} 
\end{figure}

In Fig.~\ref{fig:der-G-Q-b-P0-ink} we compare the derivative of
the redshift space distortion (green dotted line, averaged over $\mu$) with those of the growth
factor $G$ and the $Q$ variable. The $\beta$ derivative is comparable
to $G$ at $z=0$ but becomes more important at low redshifts and,
as mentioned, is similar to the $Q$ derivative in shape.

\section{Observational constraints}

\subsection{Weak lensing}

We want to investigate now the response of weak lensing (WL) to the
dark energy parameters. We proceed with a Fisher matrix as in \cite{aks},
to which we refer for the implementation details, the main difference
here being that the parameter $Q$, which was general, now has an
explicit form that describes how it changes as a function of scale
and redshift. Since $Q$ depends on $w$ and $\cs$, we can forecast
the precision with which those parameters can be extracted. We can
also try to trace where the constraints come from.

For WL experiments the important quantity is the lensing potential
which is, using the metric Eq. (\ref{pert_newton_ds}): \begin{equation}
\Phi=\phi+\psi.\end{equation}
 For a vanishing anisotropic stress the WL potential becomes: \begin{equation}
k^{2}\Phi=-2Q\frac{3H_{0}^{2}\Omega_{m,0}}{2a}\Delta_{m}.\label{eq:WL-potential}\end{equation}
 In linear perturbation theory all $k$ modes evolve independently,
so that the dark matter density contrast is often decomposed as: \begin{equation}
\Delta_{m}(a,k)=aG\left(a,k\right)\Delta_{m}\left(k\right).\label{eq:deltamatter}\end{equation}
 Here $\Delta_{m}\left(k\right)=\Delta_{m}\left(a=1,k\right)$ determines
the matter power spectrum today, $P(k)=|\Delta_{m}(k)|^{2}$ and $G\left(a,k\right)$
is the growth factor, which depends here not only time but also on
$k$ since the dark energy perturbations in general introduce a scale
dependence.

We can write Eq. (\ref{eq:WL-potential}) as: \begin{equation}
k^{2}\Phi=-3H\left(a\right)^{2}a^{3}Q\left(a,k\right)\Omega_{m}\left(a\right)G\left(a,k\right)\Delta_{m}\left(k\right)\label{eq:phiwl}\end{equation}
 where we used Eq. (\ref{eq:deltamatter}).

Hence, the lensing potential contains three conceptually different
contributions from the dark energy perturbations: 
\begin{itemize}
\item The direct contribution of the perturbations to the gravitational
potential through the factor $Q$, see section~\ref{sec:q}. 
\item The impact of the dark energy perturbations on the growth rate of
the dark matter perturbations, affecting the time dependence of $\Delta_{m}$,
through $G\left(a,k\right)$, cf section \ref{sec:gamma}. 
\item A change in the shape of the matter power spectrum $P(k)$, corresponding
to the dark energy induced $k$ dependence of $\Delta_{m}$, as discussed
in section \ref{sec:pk}. 
\end{itemize}
We consider a typical next-generation tomographic weak lensing survey
characterized by the sky fraction $f_{sky}=1/2$ and by the number of
sources per arcmin$^{2}$, $d=40$. We consider the range $10<\ell<10000$
and we extend our survey up to three different redshifts: $z_{max}=2,3,4$.
For the non linear correction we use the halo model by Smith et al.
\cite{smith-halo}. We choose as fiducial model $\Omega_{m0}=0.24$,
$h=0.7$, $\Omega_{DE}=0.737$, $\Omega_{K}=0$, $\Omega_{b}h^{2}=0.0223$,
$\tau=0.092$, $n_{s}=0.96$, $w_{0}=-0.8$, and several values for
$\cs$. The observationally borderline fiducial value of $w_{0}$
is chosen so as to maximize the impact on $Q$: values closer to $-1$
reduce the effect and therefore increase the errors on $c_{s}$.

\begin{figure}
\centering \includegraphics[width=2.6in]{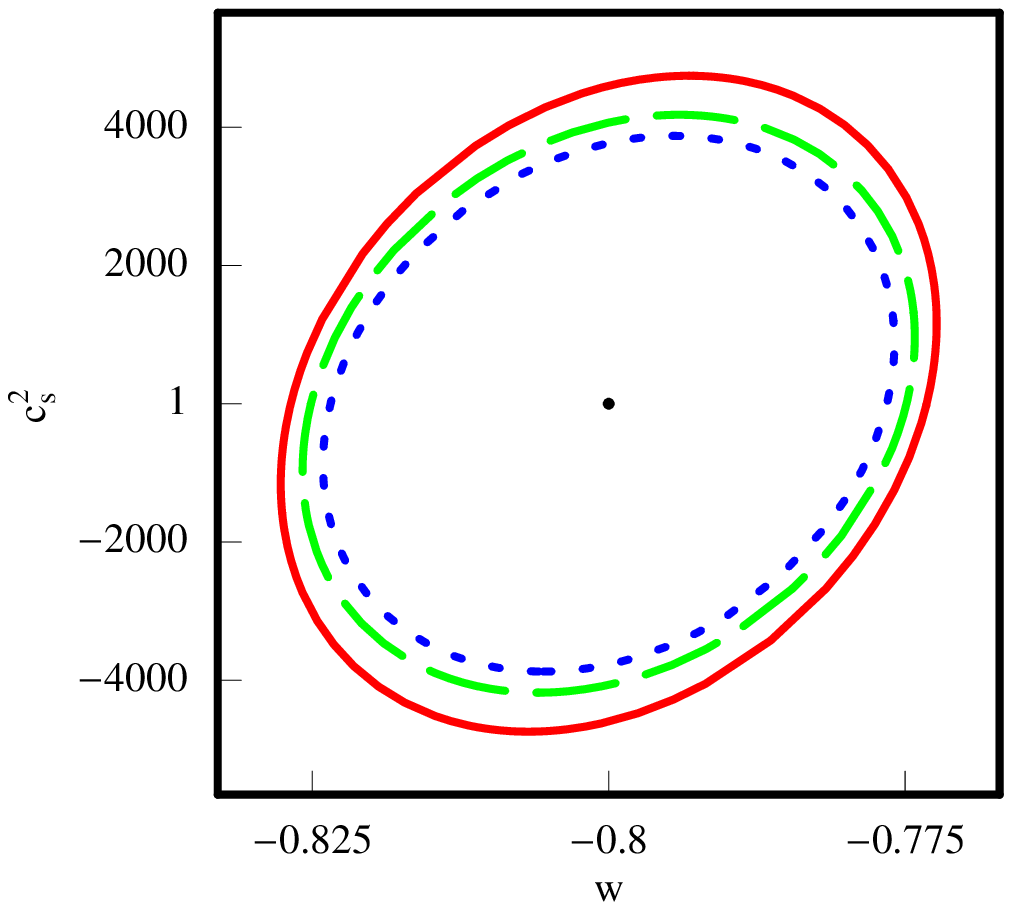}
\hspace{0.1in} \includegraphics[width=2.6in]{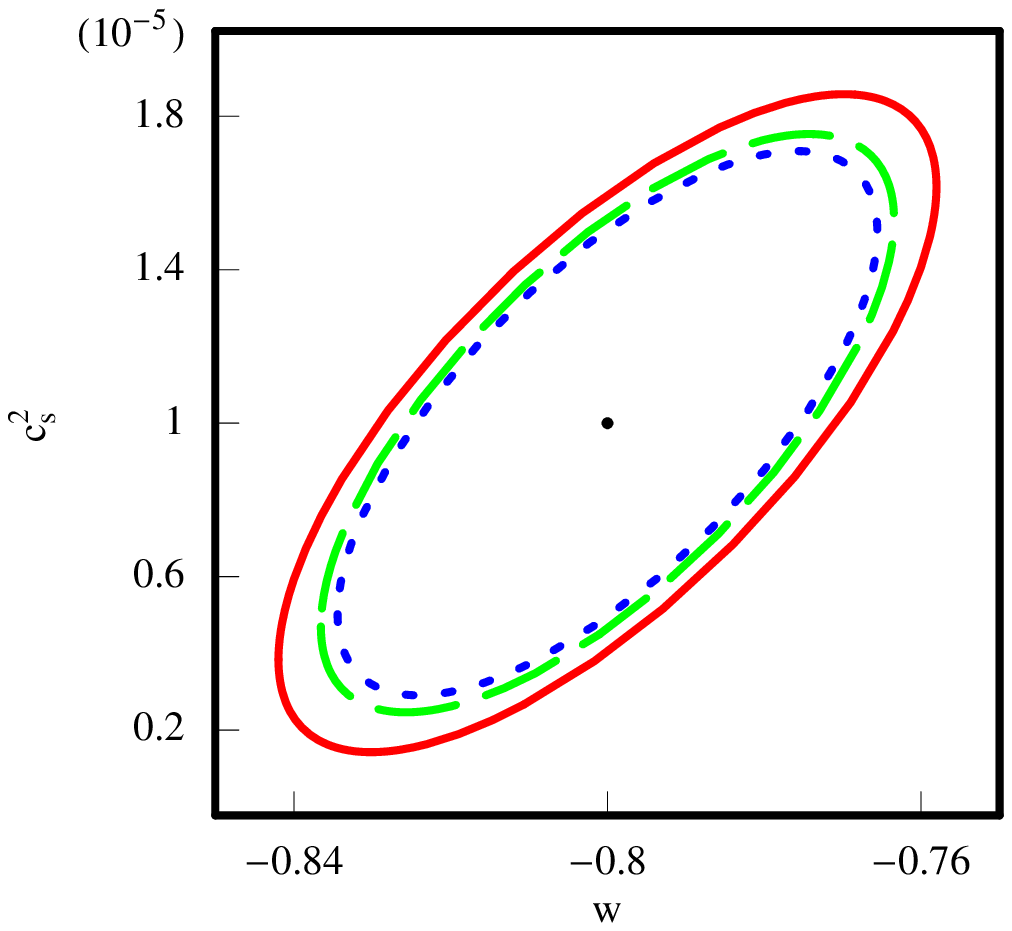}\\

\caption{Confidence region at $68\%$ for three different value of $z_{max}=2,3,4$,
blue dashed, green long-dashed and solid contour, respectively. The
upper panel shows the confidence region when the sound speed is $\cs=1$;
the bottom panel with the sound speed $\cs=10^{-5}$. The parameter
equation of state is for both cases $w_{0}=-0.8$.}

\label{fig:ellipsesw0cs} 
\end{figure}

We always plot the fully marginalized confidence ellipses (and quote
fully marginalized errors) at $68\%$, which in 2D correspond to semi-axes
of length $1.51$ time the eigenvalues. In Fig.~\ref{fig:ellipsesw0cs}
we report the confidence region for $w_{0},\cs$ for two different
values of the sound speed and $z_{max}$. For high value of the sound
speed ($\cs=1$) we find $\sigma(w_{0})=0.0171$ and the relative
error for the sound speed is $\sigma(\cs)/\cs=2768$. As expected,
WL is totally insensitive to the clustering properties of quintessence
dark energy models when the sound speed is equal to $1$. The presence
of dark energy perturbations leaves a $w$ and $\cs$ dependent signature
in the evolution of the gravitational potentials through $\Delta_{DE}/\Delta_{m}$
and, as already mentioned, the increase of the $\cs$ enhances the
suppression of dark energy perturbations which brings $Q\rightarrow1$.

Once we decrease the sound speed then dark energy perturbations are
free to grow at smaller scales. As an example we show in the lower
panel of Fig.~\ref{fig:ellipsesw0cs}, the confidence region for
$w_{0},\cs$ for $\cs=10^{-5}$, we find $\sigma(w_{0})=0.025$, $\sigma(\cs)/\cs=0.49$;
in the last case the error on the measurement on the sound speed reduced
to the 50\% of the total signal.

In Tab.~\ref{tab:errors-wl} we list the errors for $w_{0}$ and
the relative errors of $\cs$ for six different values of the sound
speed for a survey up to $z_{max}=3$.

We should note here that the Fisher matrix approach breaks down for
such large errors since it only considers the local curvature of the
likelihood at the location of the fiducial model \cite{lesgo}. If we performed
an actual analysis for data with $\cs=1$, we would find a lower limit
near $\cs=10^{-5}$ since those models are measurably different from
$\cs=1$ according to the lower panel of Figure \ref{fig:ellipsesw0cs},
and the ellipse would be cut off there. On the other hand, there would
not be any upper limit to $\cs$ since for all those cases the dark
energy perturbations are not detected.

\begin{table}
\begin{centering}
\begin{tabular}{|c|c|c|}
\hline 
\multicolumn{3}{|c|}{WL}\tabularnewline
\hline 
\hspace{0.5cm}$\cs$\hspace{0.5cm}  & $\sigma_{w_{0}}$  & $\sigma_{\cs}/\cs$ \tabularnewline
\hline 
$10^{-5}$  & $0.0257$  & $0.49$ \tabularnewline
\hline 
$10^{-4}$  & $0.0232$  & $2.48$ \tabularnewline
\hline 
$10^{-3}$  & $0.0211$  & $8.34$ \tabularnewline
\hline 
$10^{-2}$  & $0.0192$  & $44.58$ \tabularnewline
\hline 
$10^{-1}$  & $0.0183$  & $282.6$ \tabularnewline
\hline 
$1$  & $0.0171$  & $2768$ \tabularnewline
\hline
\end{tabular}
\par\end{centering}

\caption{Here are listed the errors for the equation of state parameter and
the relative errors for the sound speed at $z_{max}=3$.}

\label{tab:errors-wl} 
\end{table}

To explore more directly which term influences the most the weak lensing
signal we assume that there are two different $Q$'s: one, $Q_{\gamma}$,
which enters directly in the growth index expression (\ref{eq:gamma-Q})
and the other, $Q_{\Phi}$, which is the term that enters linearly
in the expression of the gravitational potential (\ref{eq:WL-potential}).
We concentrate here only on these terms since, as we mentioned previously,
the changes on the matter power spectrum are negligible in the interesting
range (i.e. for small sound speed). 
\begin{itemize}
\item $Q_{\gamma}$ only: Here we consider only the contribution from $Q_{\gamma}$
setting effectively $Q_{\Phi}=1$. 
\item $Q_{\Phi}$ only: Here we assume that dark energy perturbations enter
only through $Q_{\Phi}$, effectively setting the function $Q_{\gamma}=1$
in Eq. (\ref{eq:gamma-Q}). 
\end{itemize}
\begin{figure}
\centering \includegraphics[width=2.6in]{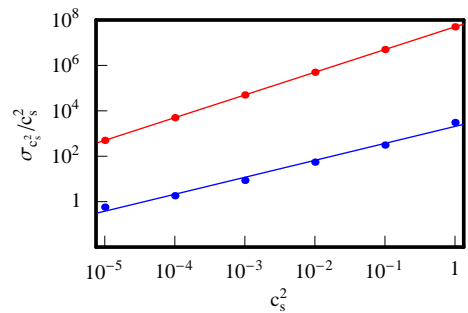}

\caption{ The figure shows the relative errors on the sound speed and their
corresponding fits for our benchmark surveys; assuming only $Q_{\gamma}$
in red and assuming only $Q_{\Phi}$ in blue.}

\label{fig:rel-errors-WL-Q-G} 
\end{figure}

We show the behavior of the errors in Fig.~\ref{fig:rel-errors-WL-Q-G}
as a function of the sound speed $\cs$. We notice that the main contribution
comes from $Q_{\Phi}$. Keeping only this term leads to the same errors
as the full expression. This is consistent with the discussion in
sections \ref{sec:q} and \ref{sec:gamma} and the derivatives shown
in Fig.~\ref{fig:der-G-Q-b-P0-ink} where we see that the growth factor
is always less sensitive to the dark energy perturbations than $Q$.
As both enter in the same way in Eq.~(\ref{eq:phiwl}) the result
is no surprise. At least for small sound speeds, gravitational lensing
constrains directly the contribution of the dark energy perturbations
to the metric, not their impact on the dark matter.

\subsection{Galaxy power spectrum}

We now explore a second probe of clustering, the galaxy power spectrum.
Following \cite{seisen} we write schematically the observed galaxy
power spectrum as: \begin{eqnarray}
P_{obs}(z,k_{r}) & = & \frac{D_{Ar}^{2}(z)H(z)}{D_{A}^{2}(z)H_{r}(z)}G^{2}(z)b(z)^{2}\left(1+\beta\mu^{2}\right)^{2}P_{0r}(k)\nonumber \\
 & + & P_{shot}(z)\end{eqnarray}
 where the subscript $r$ refers to the values assumed for the reference
(or fiducial) cosmological model, i.e. the model at which we evaluate
the Fisher matrix. Here $P_{shot}$ is the shot noise due to discreteness
in the survey, $\mu$ is the direction cosine within the survey, $P_{0r}$
is the present spectrum for the fiducial cosmology, $G(z)$ is the
linear growth factor of the matter perturbations, $b(z)$ is the bias
factor (assumed scale independent) and $D_{A}$ is the angular diameter
distance.

The wavenumber $k$ is also to be transformed between the fiducial
cosmology and the general one (\cite{seisen} and see also \cite{aqg}
and \cite{sa}, for more details). To avoid non-linearity problems
(both in the spectrum and in the bias), the Fisher matrix is calculated
up to a limiting $k_{max}(z)$ at $z$: we choose values from $0.11h/$Mpc
for low-$z$ bins to $0.3h/$Mpc for the highest $z$-bins.

Here again the galaxy power spectrum is affected by the dark energy
perturbations in three different direct ways: 
\begin{itemize}
\item through the growth factor, see Eq. (\ref{eq:gamma-Q}); 
\item through the redshift space distortions, see \ref{sec:zdist}. 
\item through the present matter power spectrum $P_{0r}$, as discussed
in \ref{sec:pk}. 
\end{itemize}
We consider a photometric survey from $z=0-2$ divided in equally
spaced bins of width $\Delta z=0.2$ as our benchmark survey; we assume
an error on the measure of redshift of about $\delta z/z=0.01$ and
an area of $20000~{\rm deg^{2}}$. These features are similar to those
of proposed experiments like JDEM and Euclid, \cite{crotts} and \cite{refregier},
respectively (see also DETF report \cite{detf}). Here too we also
consider extended surveys to $z_{max}=3$ and $z_{max}=4$. Moreover,
we assume an overall radial distribution $n\left(z\right)=z^{2}\exp[-\left(z/z_{0}\right)^{1.5}]$
where $z_{0}=z_{mean}/1.412$ and a $z_{mean}=0.9$. The bias factor
here is assumed to be only redshift dependent; we assume as fiducial
bias $b\left(z\right)=\sqrt{1+z}$. In general, however, dark energy
perturbations could lead to a scale dependence of the bias factor.
Our fiducial model is exactly the same as for the WL survey considered
in the previous section.

\begin{figure}
\centering \includegraphics[width=2.6in]{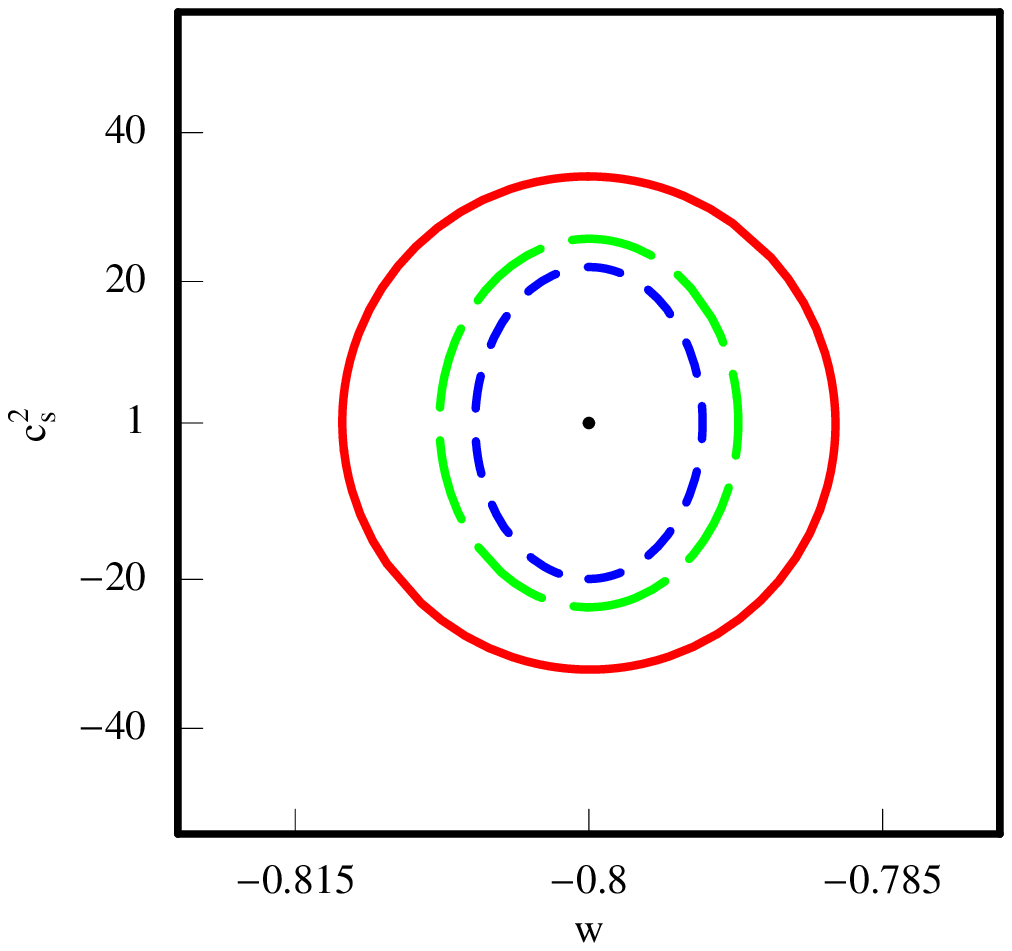}
\hspace{0.1in} \includegraphics[width=2.8in]{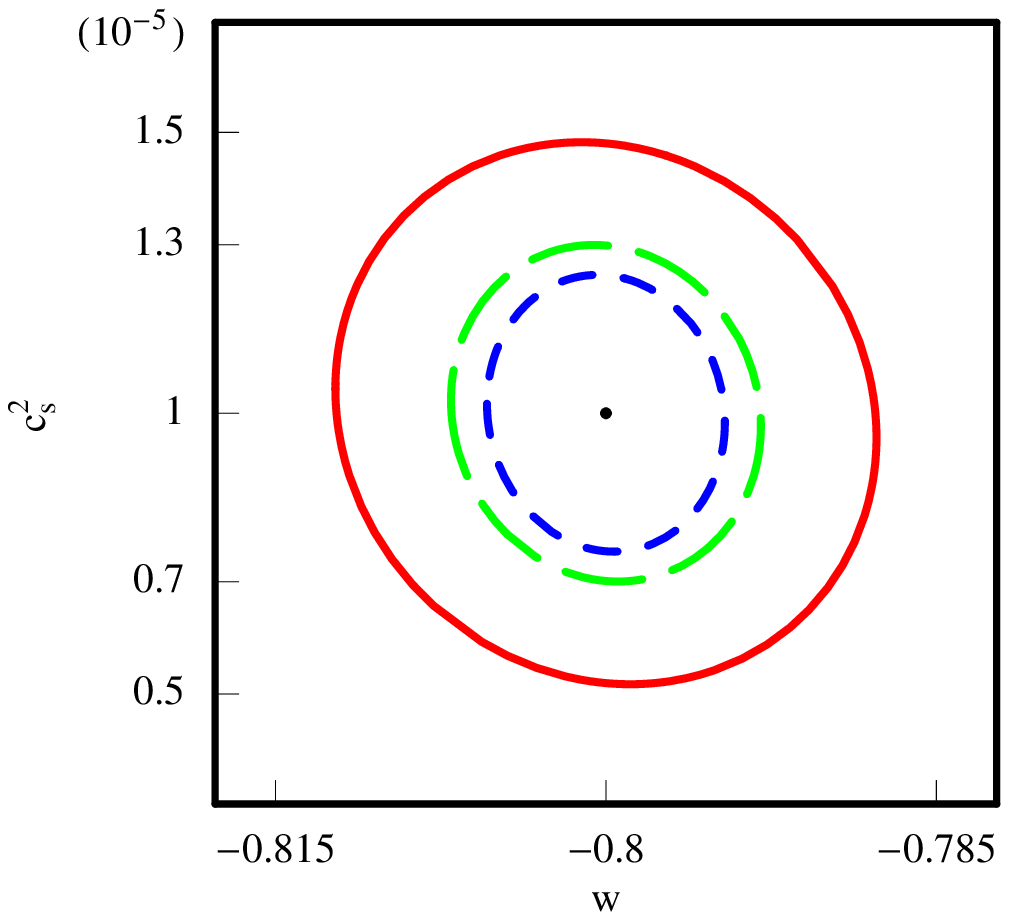}\\

\caption{ Confidence region at $68\%$ for three different value of $z_{max}=2,3,4$,
blue dashed, green long-dashed and solid counter, respectively. The
upper panel shows the confidence region when the sound speed is $\cs=1$;
the bottom panel with the sound speed $\cs=10^{-5}$. The parameter
equation of state is for both cases $w_{0}=-0.8$.}

\label{fig:ellipsesw0cs-pk} 
\end{figure}

In Fig.~\ref{fig:ellipsesw0cs-pk} we report the confidence region
for $w_{0},\cs$ for two different values of the sound speed and $z_{max}$.
For high values of the sound speed ($\cs=1$) we find, for our benchmark
survey: $\sigma(w_{0})=0.0088$, and $\sigma(\cs)/\cs=22.07$. Here
again we find that galaxy power spectrum is not sensitive to the clustering
properties of dark energy when the sound speed is of order unity.
If we decrease the sound speed down to $\cs=10^{-5}$ then the errors
are $\sigma(w_{0})=0.0091$, $\sigma(\cs)/\cs=0.32$.

In Tab.~\ref{tab:errors-BAO} we listed explicitly the errors for
$w_{0}$ and the relative errors of $\cs$ for six different values
of the sound speed for a benchmark survey up to $z_{max}=2$.

\begin{table}
\begin{centering}
\begin{tabular}{|c|c|c|}
\hline 
\multicolumn{3}{|c|}{$P\lkr$}\tabularnewline
\hline 
\hspace{0.5cm}$\cs$\hspace{0.5cm}  & $\sigma_{w_{0}}$  & $\sigma_{\cs}/\cs$ \tabularnewline
\hline 
$10^{-5}$  & $0.009157$  & $0.32$ \tabularnewline
\hline 
$10^{-4}$  & $0.009121$  & $0.66$ \tabularnewline
\hline 
$10^{-3}$  & $0.009112$  & $1.41$ \tabularnewline
\hline 
$10^{-2}$  & $0.009069$  & $2.75$ \tabularnewline
\hline 
$10^{-1}$  & $0.008960$  & $14.91$ \tabularnewline
\hline 
$1$  & $0.008825$  & $22.07$ \tabularnewline
\hline
\end{tabular}
\par\end{centering}

\caption{Errors for the equation of state parameter and the relative errors
for the sound speed for $z_{max}=2$.}

\label{tab:errors-BAO} 
\end{table}

\begin{figure}
\centering \includegraphics[width=2.8in]{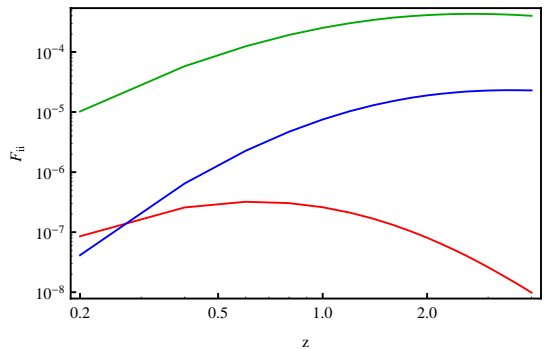}

\caption{The figures shows the Fisher matrix element $G-G$, $\beta-\beta$,
$P_{0}-P_{0}$, solid blue, red and green lines, respectively. For
$\cs=1$}

\label{fig-fisher-cs1} 
\end{figure}

\begin{figure}
\centering \includegraphics[width=2.8in]{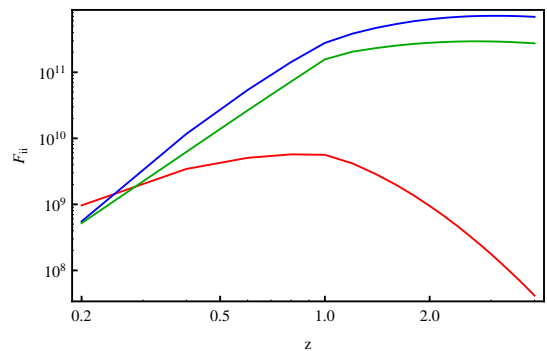}

\caption{The upper panel shows $G-G$, $\beta-\beta$, $P_{0}-P_{0}$, solid
blue, red and green lines, respectively. For $\cs=10^{-5}$}

\label{fig-fisher-cs000001} 
\end{figure}

We can ask again the question what part of the power spectrum is most
sensitive to the dark energy perturbations, $G$, $\beta$ or $P_{0}(k)$?
In order to disentangle their relative contributions, we plot the
corresponding diagonal Fisher matrix terms containing the sound speed
as a function of redshift. The results are shown in Fig.~\ref{fig-fisher-cs1}
for $\cs=1$ and in Fig. ~\ref{fig-fisher-cs000001} for $\cs=10^{-5}$.
For large sound speed the shape of the dark matter power spectrum
is formally the most sensitive, but not sufficiently to constrain
the perturbations. In the more interesting case where the sound speed
is lower and the perturbations can be detected, it is instead the
growth factor that is the most sensitive.

In both cases the contribution from redshift space distortions to
the overall constraints is sub-dominant except at redshifts below
about $z=0.3$ and for small $\cs$, cf Figures \ref{fig-fisher-cs1}
and \ref{fig-fisher-cs000001}. In the low-redshift region the growth
factor is less sensitive as it is an integral from $z=0$. To further
clarify the importance of the redshift space distortions, we performed
our analysis excluding the dependence of dark energy perturbations
on the redshift distortion parameter. The result, given in Tab.~\ref{tab:errors-W-BAO-nobeta},
shows that for small sound speeds the contribution from $\beta$ is
non-negligible when we compare it to Table \ref{tab:errors-BAO}.

\subsection{The scaling of the errors with $\cs$}

\begin{figure}
\centering \includegraphics[width=2.8in]{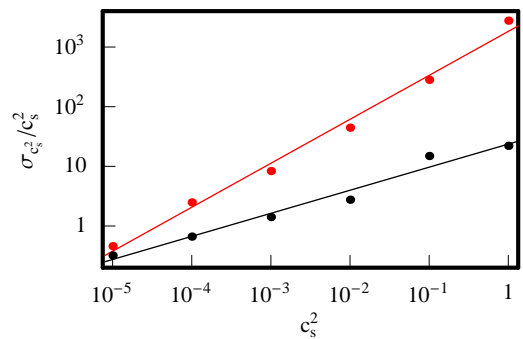}

\caption{The figure shows the relative errors on the sound speed for two different
experiments and their fits: WL in red and $P(k)$ in black, both for
our benchmark survey.}

\label{fig:errors} 
\end{figure}

In Fig.~\ref{fig:errors} the behavior in logarithmic scale of the
sound speed errors $\sigma(\cs)/\cs$ are shown as a function of the
sound speed itself. We notice that the errors appear to scale as a
power law for both probes, over the range of sound speeds given. For
the fiducial surveys we find \begin{equation}
\frac{\sigma_{\cs}}{\cs}=1830\left(\cs\right)^{0.74}\end{equation}
 for the weak lensing and\begin{equation}
\frac{\sigma_{\cs}}{\cs}=20.6\left(\cs\right)^{0.387}\end{equation}
 for the power spectrum.

\begin{table}
\begin{centering}
\begin{tabular}{|c|c|c|c|}
\hline 
\multicolumn{4}{|c|}{$P\lkr$}\tabularnewline
\hline 
 & $z_{max}=2$  & $z_{max}=3$  & $z_{max}=4$ \tabularnewline
\hline 
\hspace{0.5cm}$\cs$\hspace{0.5cm}  & $\sigma_{\cs}/\cs$  & $\sigma_{\cs}/\cs$  & $\sigma_{\cs}/\cs$ \tabularnewline
\hline 
$10^{-5}$  & $0.67$  & $0.48$  & $0.41$\tabularnewline
\hline 
$1$  & $22.48$  & $15.32$  & $12.69$\tabularnewline
\hline
\end{tabular}
\par\end{centering}

\caption{Relative errors for $\cs$ for $P\left(k\right)$ for different $z_{max}$
without redshift distortion.}

\label{tab:errors-W-BAO-nobeta} 
\end{table}

\subsection{Combined Fisher matrices and degeneracies.}

We can also combine the two probes, the galaxy power spectrum and
weak lensing, to improve the sensitivity to the dark energy perturbations,
cf. Tab.~\ref{tab:combined-BAO-WL}. As expected the errors decrease
by about $30\%$ for low values of the sound speed and remain unaltered
when the sound speed is close to unity.

We find that the sound speed does not suffer of any strong degeneracies
with the other parameters considered in this work. We evaluated the
dimensionless correlation matrix: \begin{equation}
{\normalcolor \mathrm{Corr}}_{ij}=\frac{C_{ij}}{\sqrt{C_{ii}C_{jj}}}\end{equation}
 where $C_{ij}$ is the inverse of the Fisher Matrix $F_{ij}$. For
instance, for $\cs=10^{-5}$, we find for all parameters a correlation
factor with $\cs$ of at most $0.3$ for galaxy power spectrum experiments
and at most $0.23$ for WL experiments; however, for the latter case we find  
for $\Omega_{m,0}-\cs$ terms a correlation value of about $0.68$. Increasing
the sound speed the correlation terms decrease for galaxy power spectrum
case and WL case; for the latter only the term $\Omega_{b}h^{2}-\cs$
increases up to $0.78$, for $\cs=1$.

%
\begin{table}
\begin{centering}
\begin{tabular}{|c|c|c|c|c|c|c|}
\hline 
 & $\Omega_{m,0}h^{2}$  & $\Omega_{b,0}h^{2}$  & $n_{s}$  & $\Omega_{m,0}$  & $w_{0}$  & $\cs$ \tabularnewline
\hline 
$\Omega_{m,0}h^{2}$  & $1$  & $-0.57$  & $-0.55$  & $-0.97$  & $-0.57$  & $-0.27$\tabularnewline
\hline 
$\Omega_{b,0}h^{2}$  & $-0.57$  & $1$  & $0.23$  & $0.64$  & $0.49$  & $0.22$\tabularnewline
\hline 
$n_{s}$  & $-0.55$  & $0.23$  & $1$  & $0.43$  & $0.57$  & $0.18$\tabularnewline
\hline 
$\Omega_{m,0}$  & $-0.97$  & $0.64$  & $0.43$  & $1$  & $0.46$  & $0.25$\tabularnewline
\hline 
$w_{0}$  & $-0.57$  & $0.49$  & $0.57$  & $0.46$  & $1$  & $0.30$\tabularnewline
\hline 
$\cs$  & $-0.27$  & $0.22$  & $0.18$  & $0.25$  & $0.30$  & $1$\tabularnewline
\hline
\end{tabular}
\par\end{centering}

\caption{Correlation matrix for galaxy power spectrum experiments and $\cs=10^{-5}$.}

\label{tab:Deg-bao} 
\end{table}


%
\begin{table}
\begin{centering}
\begin{tabular}{|c|c|c|c|c|c|c|}
\hline 
 & $\Omega_{m,0}h^{2}$  & $\Omega_{b,0}h^{2}$  & $n_{s}$  & $\Omega_{m,0}$  & $w_{0}$  & $\cs$ \tabularnewline
\hline 
$\Omega_{m,0}h^{2}$  & $1$  & $-0.25$  & $0.88$  & $-0.09$  & $0.17$  & $-0.01$\tabularnewline
\hline 
$\Omega_{b,0}h^{2}$  & $-0.25$  & $1$  & $0.09$  & $0.019$  & $-0.39$  & $-0.18$\tabularnewline
\hline 
$n_{s}$  & $0.88$  & $0.09$  & $1$  & $-0.32$  & $0.22$  & $-0.16$\tabularnewline
\hline 
$\Omega_{m,0}$  & $-0.09$  & $0.19$  & $-0.32$  & $1$  & $-0.86$  & $0.68$\tabularnewline
\hline 
$w_{0}$  & $0.17$  & $-0.39$  & $0.22$  & $-0.86$  & $1$  & $-0.23$\tabularnewline
\hline 
$\cs$  & $-0.01$  & $-0.18$  & $-0.16$  & $0.68$  & $-0.23$  & $1$\tabularnewline
\hline
\end{tabular}
\par\end{centering}

\caption{Correlation matrix for WL experiments and $\cs=10^{-5}$.}

\label{tab:Deg-wl} 
\end{table}


\subsection{An integrated measure of dark energy clustering}

So far we have been studying the errors on the dark energy parameters
$w$ and $\cs$. A subtly different question is whether we are
able to detect dark energy perturbations or not -- a question not
about parameter constraints but about model probabilities \cite{mpclk}.
It is not straightforward to use our constraints e.g. on the sound
speed for this purpose, as it does not make sense to consider the
speed of sound if there are no perturbations at all, i.e. a model
without perturbations is not nested in the quintessence model considered
here. This is just another way to say that all quintessence models
(except those for which $w=-1$, equivalent to a cosmological constant) necessarily
have perturbations.

But from the errors on $w$ and $\cs$ we can forecast the constraints
on the amount of deviation of the Poisson equation from the expectation
due to the dark matter and other known contributions, given by $Q$.
A model without dark energy perturbations would then be characterized
by $Q=1$. However, $Q$ is a function which evolves and additionally
can have a scale dependence. It is then useful to define a compact
way to express the deviation integrated over $z$ and $k$. We define
therefore the average quantity \begin{equation}
W(\Omega_{m},w,c_{s}^{2})\equiv\frac{4\pi}{V_{k}\Delta z}\int\left|Q(k,z)-1\right|dzk^{2}dk\end{equation}
 where $\Delta z$ is the observed redshift range and $V_{k}=4\pi k_{max}^{2}/3$
is the momentum volume. Finding a deviation of $W$ from zero would
signal some clustering of dark energy. We note that it is straightforward
to focus on certain scales or epochs by introducing an appropriate
weight function into the integral above.

In order to obtain constraints on $W$ we need to marginalize the
Fisher matrix over all parameters except $p_{i}=\{\Omega_{m},w_{0},c_{s}^{2}\}$
and then project it over the new parameter set $q_{j}=\{W,w_{0},\cs\}$.
Defining $J_{ij}=(\partial p_{i}/\partial q_{j})_{r}$ we have \begin{equation}
\sigma_{q_{ii}}^{2}=(J_{ik}F_{kl}J_{li})^{-1}.\end{equation}
 The errors on $W$ for the $P(k)$ case are reported in Tab.~\ref{tab:errors-W-Pk}
and for WL  in Tab.~\ref{tab:errors-W-WL}. Since the
derivative of $Q$ with respect the sound speed contains a term which
goes like $1/\cs$, the dominating source of error on $W$ comes from
the errors on $c_{s}^{2}$, as long as $\cs\ll1$. Therefore a quick
estimate can be obtained by approximating \begin{eqnarray}
 &  & \frac{\sigma_{W}}{W}=\left|\frac{\partial\ln W}{\partial\ln\cs}\right|\frac{\sigma_{\cs}}{\cs}\approx\frac{\sigma_{\cs}}{\cs}\end{eqnarray}
 and we find that in the present case the relative errors on $W$ are
similar to those on $\cs$. This supports the naive interpretation
that we can only detect the perturbations if we can measure the sound
speed, i.e. if the relative error on $\cs$ is less than unity.

However, we are now in a position where we can use model comparison
techniques to decide whether $W\neq0$ has been detected or not. Using
the Savage-Dickey density ratio, we find that the relative model probability
is given by the value of the normalized posterior (marginalized over
all common parameters, i.e. all parameters except $W$) at $W=0$,
divided by the value of the normalized prior at the same place. To
simplify the calculations, we choose a uniform prior in all variables
except $W$, and for that variable a Gaussian pdf centered at $W=0$
with a width (variance) $\Sigma^{2}$. This then leads to a Bayes
factor of \begin{equation}
B_{W}=\sqrt{\frac{\bar{\sigma}^{2}}{\Sigma^{2}}}\exp\left\{ \frac{\bar{W}^{2}}{2\bar{\sigma}^{2}}\right\} .\end{equation}
 Here $\bar{W}$ is the value on which the posterior for $W$ is centered,
and $\bar{\sigma}^{2}$ the variance of $W$. Given the fiducial value
$W_{{\rm fid}}$ and the Fisher matrix error $\sigma_{W}^{2}$, which
characterize the likelihood, we find for the posterior the shifted
values (due to the prior, see e.g. the appendix of Ref.~\cite{ktp})
\begin{eqnarray}
\bar{\sigma}^{2} & = & \frac{\Sigma^{2}\sigma_{W}^{2}}{\Sigma^{2}+\sigma_{W}^{2}}\\
\bar{W} & = & \frac{\Sigma^{2}}{\Sigma^{2}+\sigma_{W}^{2}}W_{{\rm fid}}.\end{eqnarray}
 However, it is not obvious how we should choose the width of the
prior, for a general model with unknown sound speed. One possibility
is to use the most optimistic value, for which the model discrimination
is maximal \cite{gt}. To find this value, we maximize $B_{W}$ with
respect to $\Sigma$, which leads to $\Sigma^{2}=W_{{\rm fid}}^{2}-\sigma_{W}^{2}$.
Using this prescription (valid only for $W_{{\rm fid}}>\sigma_{W}$)
we finally obtain \begin{equation}
B_{W}=\sqrt{\frac{\sigma_{W}^{2}}{W_{{\rm fid}}^{2}}}\exp\left\{ \frac{1}{2}\left(\frac{W_{{\rm fid}}^{2}}{\sigma_{W}^{2}}-1\right)\right\} .\end{equation}
 This formula depends only on the ratio $W_{{\rm fid}}/\sigma_{W}$,
and if we demand strong evidence in favor of the presence of perturbations,
$\ln B>5$, we find that we need $\sigma_{W}/W_{{\rm fid}}\lesssim0.27$,
i.e. not quite a 4-$\sigma$ detection. We remind the reader that
this corresponds to the choice of prior that maximally favors the
detection of perturbations, any other choice would need a stronger
detection of $W$ to reach the same Bayes factor.

Looking at tables \ref{tab:errors-W-Pk}, \ref{tab:errors-W-WL} and
\ref{tab:combined-BAO-WL} we can see that in our class of quintessence
models and for the observations that we consider here, we can only
hope to strongly favor the presence of dark energy perturbations if
the sound speed is $c_{s}<0.01$.

\begin{table}
\begin{centering}
\begin{tabular}{|c|c|c|}
\hline 
\multicolumn{3}{|c|}{$P(k)$}\tabularnewline
\hline 
\hspace{0.5cm}$\cs$\hspace{0.5cm}  & $\sigma_{W}/W$  & $\sigma_{\cs}/\cs$ \tabularnewline
\hline 
$10^{-5}$  & $0.15$  & $0.32$ \tabularnewline
\hline 
$10^{-4}$  & $0.42$  & $0.66$ \tabularnewline
\hline 
$10^{-3}$  & $1.16$  & $1.41$ \tabularnewline
\hline 
$10^{-2}$  & $2.53$  & $2.79$ \tabularnewline
\hline 
$10^{-1}$  & $13.75$  & $14.91$ \tabularnewline
\hline 
$1$  & $21.46$  & $22.07$ \tabularnewline
\hline
\end{tabular}
\par\end{centering}

\caption{Relative errors for $W$ and the corresponding $\cs$ for $P(k)$.}

\label{tab:errors-W-Pk} 
\end{table}

\begin{table}
\begin{centering}
\begin{tabular}{|c|c|c|}
\hline 
\multicolumn{3}{|c|}{$WL$}\tabularnewline
\hline 
\hspace{0.5cm}$\cs$\hspace{0.5cm}  & $\sigma_{W}/W$  & $\sigma_{\cs}/\cs$ \tabularnewline
\hline 
$10^{-5}$  & $0.31$  & $0.49$ \tabularnewline
\hline 
$10^{-4}$  & $2.13$  & $2.48$ \tabularnewline
\hline 
$10^{-3}$  & $8.26$  & $8.34$ \tabularnewline
\hline 
$10^{-2}$  & $44.39$  & $44.58$ \tabularnewline
\hline 
$10^{-1}$  & $281.7$  & $282.6$ \tabularnewline
\hline 
$1$  & $2766$  & $2768$ \tabularnewline
\hline
\end{tabular}
\par\end{centering}

\caption{Relative errors for $W$ and the corresponding $\cs$ for WL.}

\label{tab:errors-W-WL} 
\end{table}

\begin{table}
\begin{centering}
\begin{tabular}{|c|c|c|c|}
\hline 
\multicolumn{4}{|c|}{$P\lkr+$WL}\tabularnewline
\hline 
\hspace{0.5cm}$\cs$\hspace{0.5cm}  & $\sigma_{w_{0}}$  & $\sigma_{\cs}/\cs$  & $\sigma_{W}/W$\tabularnewline
\hline 
$10^{-5}$  & $0.00639$  & $0.15$  & $0.11$\tabularnewline
\hline 
$10^{-4}$  & $0.00581$  & $0.41$  & $0.36$ \tabularnewline
\hline 
$10^{-3}$  & $0.00547$  & $0.87$  & $1.02$ \tabularnewline
\hline 
$10^{-2}$  & $0.00531$  & $2.48$  & $2.39$ \tabularnewline
\hline 
$10^{-1}$  & $0.00528$  & $14.79$  & $13.14$\tabularnewline
\hline 
$1$  & $0.00524$  & $22.05$  & $21.29$ \tabularnewline
\hline
\end{tabular}
\par\end{centering}

\caption{Errors for the equation of state parameter, the relative errors for
the sound speed and the relative errors for the variable $W$ combining
WL and galaxy power spectrum probes.}

\label{tab:combined-BAO-WL} 
\end{table}

\section{Conclusion}

In this paper we extended the idea proposed in \cite{sk}. We investigated
the effect of a generic modification of Poisson equation $Q(k,a)$
in the galaxy power spectrum and in the weak lensing convergence power
spectrum induced solely by dark energy clustering. In other words,
we are using perturbations in the metric and in the dark matter to
detect the tiny traces of dark energy perturbations.

We identified several effects: on the power spectrum growth, on its
shape, and (for the galaxy clustering probe) on the redshift distortion.
These effect are completely characterized by $w$ and $c_{s}^{2}$.
We performed a Fisher matrix analysis of future large-scale surveys
to forecast the errors on $w$ and $c_{s}^{2}$ and therefore on $Q(k,a)$
or on its average version, $W$. As perhaps expected, we find that
dark energy perturbations have a very small effect on dark matter
clustering unless the sound speed is extremely small, $c_{s}\le0.01$.
For $c_{s}\le0.01$, we find that $W$ (that is, the average deviation
for a standard Poisson equation) could be constrained to within 66\%
or better. Let us remind the reader that in order to boost the observable
effect, we always assumed $w=-0.8$: for values closer to $-1$ the
sensitivity to $\cs$ is further reduced; as a test we performed the
calculation for $w=-0.9$ and $\cs=10^{-5}$ and we find $\sigma_{\cs}/\cs=2.6$
and $\sigma_{\cs}/\cs=1.09$ for WL and galaxy power spectrum experiments,
respectively.

Such small sound speeds are not in contrast with the fundamental expectation
of dark energy being much smoother that dark matter: even with $c_{s}\approx0.01$,
dark energy perturbations are more than one order of magnitude weaker
than dark matter ones (at least for the class of models investigated
here) and safely below non-linearity at the present time at all scales.
Models of {}``cold'' dark energy with $\cs=0$ are interesting because they can
cross the phantom divide \cite{ksphantom} and contribute to the cluster
masses \cite{vernizzi}. Small $c_{s}$ could be constructed for instance
with scalar fields with non-standard kinetic energy terms. Here we
showed that future large scale surveys have the potential to rule
out or confirm this class of models. As an example we report here
the absolute errors for the sound speed $\cs=0$: $\sigma_{\cs}=0.344 \cdot 10^{-9}$
and $\sigma_{\cs}=0.211\cdot 10^{-7}$ for WL and galaxy power spectrum
experiments respectively.


\begin{acknowledgments}

D. S. acknowledges support by the Spanish ${\rm MICINN}$ under the project 
${\rm AYA}2009-13936-{\rm C}06-06$ and the EU FP6 Marie Curie Research and Training Network 
"UniverseNet" (${\rm MRTN-CT}-2006-035863$).
M.K. acknowledges support by the Swiss NSF.

\end{acknowledgments}


\begin{thebibliography}{34}
\bibitem{sn1} A. G. Riess et al., Astronomical J. \textbf{116}, 1009
(1998).

\bibitem{sn2} S. Perlmutter et al., Astrophys. J. \textbf{517}, 565
(1999).

\bibitem{iklf} S. Ilic, M. Kunz, A.R. Liddle and J.A. Frieman, 
Phys. Rev. D \textbf{81}, 103502 (2010).

\bibitem{cald}R. Caldwell, R. Dave, and P. J. Steinhardt, Phys. Rev.
Lett. 80 1582 (1998).

\bibitem{ksphantom} M. Kunz and D. Sapone, Phys. Rev. D \textbf{74},
123503 (2006).


\bibitem{mfb} V.F. Mukhanov, H.A. Feldman and R.H. Brandenberger,
Phys. Rep. \textbf{215}, 206 (1992).

\bibitem{abp} V. Acquaviva, C. Baccigalupi and F. Perrotta, Phys.
Rev. D \textbf{70}, 023515 (2004).

\bibitem{lss} A. Lue, R. Scoccimarro and G.D. Starkmann, Phys. Rev.
D \textbf{69}, 124015 (2004).

\bibitem{km} K. Koyama and R. Maartens, JCAP \textbf{0601}, 016 (2006).

\bibitem{ks2} M. Kunz and D. Sapone, Phys. Rev. Lett. \textbf{98},
121301 (2007).

\bibitem{hs} W. Hu and I. Sawicki, Phys. Rev. D \textbf{76}, 104043
(2007).

\bibitem{aks} L. Amendola, M. Kunz and D. Sapone, JCAP {\textbf{0}4},
013 (2008).

\bibitem{sk} D. Sapone and M. Kunz, Phys. Rev D {\textbf{80}},
083519 (2009).

\bibitem{kess}J. K. Erickson, J R. R. Caldwell, P. J. Steinhardt,
C. Armendariz-Picon, \& V. Mukhanov, Phys. Rev. Lett. \textbf{88},
121301 (2002).

\bibitem{vernizzi} P. Creminelli, G. D'Amico, J. Norena, L. Senatore,
\& F. Vernizzi, JCAP \textbf{3} (2010), 27

\bibitem{gklp} C. Gao, M. Kunz, A.R. Liddle and D. Parkinson, Phys.
Rev. D \textbf{81}, 043520 (2010).

\bibitem{dark_degen} M. Kunz, Phys. Rev. D \textbf{80} 123001 (2009).


\bibitem{wellew} J. Weller and A. M. Lewis, Mon. Not. Roy. Astron.
Soc. \textbf{346} 987 (2003).

\bibitem{beandore} R. Bean and O. Dore Phys. Rev. D \textbf{69},
083503 (2004).

\bibitem{detf} A. Albrecht et al., Dark Energy Task Force report
to the Astronomy and Astrophysics Advisory Committee, DETF \url{http://www.nsf.gov/mps/ast/detf.jsp}

\bibitem{deput} R. de Putter, D. Huterer and E.V. Linder,
Phys. Rev. D {\bf 81}, 103513 (2010).

\bibitem{lica} E.V. Linder and R.N. Cahn, Astropart. Phys. \textbf{28},
481 (2007).

\bibitem{camb} A. Lewis, A. Challinor and A. Lasenby, Astrophys.
J. \textbf{538}, 473 (2000).

\bibitem{ehu}D. Eisenstein and W. Hu, Astrophys. J. \textbf{511},
5 (1999).

\bibitem{flow1} Courteau, S.A., Strauss, M. A., \& Willick, J. A.,
Eds., 2000, ASP Conf. Ser. 201, Cosmic Flows (San Francisco: ASP)

\bibitem{flow2} P.G. Ferreira, R. Juszkiewicz, H.A. Feldman, M. Davis,
A.H. Jaffe, Astrophys. J. \textbf{515} L1 (1999).

\bibitem{smith-halo} R.~E.~Smith et al., MNRAS \textbf{341}, 1311
(2003).

\bibitem{lesgo} G. Ballesteros and J. Lesgourgues, arXiv:1004.5509 (2010).

\bibitem{seisen} H. J. Seo and D. J. Eisenstein, Ap. J. \textbf{598},
720 (2003).

\bibitem{aqg} L. Amendola, C. Quercellini and E. Giallongo, MNRAS
\textbf{357}, 429 (2005).

\bibitem{sa} D. Sapone and L. Amendola, astro-ph/0709.2792.

\bibitem{crotts} A. Crotts at al. (2005), astro-ph/0507043.

\bibitem{refregier} A. Refregier et al. (2006); Procs. of symposium
{}``Astronomical Telescopes and Instrumentation''; astro-ph/0610062.

\bibitem{mpclk} P. Mukherjee, D. Parkinson, P.S. Corasaniti, A.R.
Liddle and M. Kunz, Mon. Not. Royal Astron. Soc. \textbf{369}, 1725
(2006).


\bibitem{ktp} M. Kunz, R. Trotta and D. Parkinson, Phys. Rev. D \textbf{74},
023503 (2006).

\bibitem{gt} C. Gordon and R. Trotta, Mon. Not. Roy. Astron. Soc.
\textbf{382}, 1859 (2007).


\end{thebibliography}
\end{document}